\documentclass[draft]{aipproc}

\layoutstyle{6x9}

\begin{document}

\title{Super-Dense Matter at Super- Strong Magnetic Fields}


\author{Efrain J. Ferrer and Vivian de la Incera}{
  address={Department of Physics, Western Illinois University,
Macomb, IL 61455, USA} }


\begin{abstract}
 Our Universe is full of regions where extreme physical conditions are realized.
A most intriguing case is the super-dense core of neutron stars,
some of which also have super-strong magnetic fields, hence called
magnetars. In this paper we review the current understanding of the
physical properties  of the different phases of quark matter at very
high densities in the presence of large magnetic fields. We also
discuss how the Meissner instability produced at moderate densities
by a pairing stress due to the medium neutrality and
$\beta$-equilibrium constraints can lead to the spontaneous
generation of a magnetic field.
\end{abstract}

\maketitle


\section{Introduction}

Neutron stars, the leftover core of an exploding supergiant star,
are so compact that gravity binds nucleons there 10 times more
strongly than the strong-nuclear force binds nucleons in nuclei.
Their average density can reach up to $10^{18} kg/m^3$. At those
tremendous densities the superdense matter at the cores of those
compact objects can realize a quark deconfined phase that produces a
color superconducting state \cite{CS, CFL}.

A common characteristic of neutron stars is their strong
magnetization. It is known that, since the surface area of a
supergiant star shrinks by a factor of about $10^{10}$ after the
supernova collapse, the resultant field strength is increased by the
same factor. Surface magnetic fields of these compact objects range
from $B=1.7\times 10^{8} G$ (PSR B1957+20) up to $2.1\times 10^{13}
G$ (PSR B0154+61), with a typical value of $10^{12} G$
\cite{Taylor}. The existence of stellar objects - known as
magnetars- with even stronger surface magnetic fields of order
$B\sim 10^{14}-10^{15} G$ \cite{magnetars} is already an
observational fact. In the core of these compact objects, the field
may be considerably larger due to flux conservation during the core
collapse. By applying the scalar virial theorem it can be shown that
the interior field can reach values of order $B \sim 10^{18} G$
\cite{Lai}.

From the above arguments, it becomes imperative to investigate the
interplay between color superconductivity and strong magnetic fields
in order to understand the real state of matter in the inner region
of compact astronomical objects. Although a color superconductor
(CS) is in principle an electric superconductor because the diquark
condensate carries nonzero electric charge, in spin-zero phases like
the color-flavor-locked (CFL) and the 2SC phases
\cite{alf-raj-wil-99/537}, there is no Meissner effect for a new
in-medium electromagnetic field $\widetilde{A}_{\mu}$. This
in-medium or "rotated" electromagnetic field is a combination of the
regular electromagnetic field and the $8^{th}$ gluon
\cite{alf-raj-wil-99/537}. As the quark pairs are all neutral with
respect to the "rotated" electromagnetic charge $\widetilde{Q}$, the
"rotated" electromagnetic field $\widetilde{A}_{\mu}$ remains
long-range within the superconductor.

It has been recently found \cite{MCFL}-\cite{PCFL}, that the
magnetic field can modify the ground state of the CS. An external
magnetic field can lead to the splitting of the gap parameters
\cite{MCFL} and produce Haas-van Alphen oscillations of the gap
\cite{Warringa} and the magnetization \cite{Igor}. Moreover, it has
been shown that the appearance of chromomagnetic instabilities
triggered either by a strong magnetic field \cite{PCFL} or, in the
absence of the field, by pairing stress occurring at moderate
densities, leads to the formation of inhomogeneous gluon condensates
that can either boost the existing field \cite{PCFL} or
spontaneously generate one \cite{Vortex}.

These effects, all of which connect magnetism with CS, create a new
scenario for magnetized high-dense matter where different phases can
be realized. In that scenario, the boundaries between different
magnetic phases are determined by the various scales that
characterize the CS \cite{Phases}. Ignoring the quark masses, the
main scales of this medium are the baryon chemical potential $\mu$,
the dynamically generated gluon mass $m_{M}\sim g\mu$ and the gap
parameter $\Delta\sim \frac{\mu}{g^{5}} e^{-\alpha/g}$, with
$\alpha$ a constant that is dominated by magnetic gluon exchanges
\cite{Son-2}. We can assume that at sufficiently large $\mu$, the
running strong coupling $g$ becomes weak enough for the hierarchy of
the scales to be $\Delta\ll m_{g}\ll \mu$. In a CS with three
flavors a phase transmutation from CFL to the so-called magnetic CFL
phase (MCFL) takes place when $\widetilde{B} \simeq
\Delta^{2}_{CFL}$ \cite{Phases}. During the phase transmutation no
symmetry breaking occurs, since in principle once a magnetic field
is present the symmetry is theoretically that of the MCFL phase. For
even larger fields ($\widetilde{B} \geq \widetilde{B}_{PCFL} =
m_{M}^2$, with $m_{M}$ being the magnetic mass of the charged
gluons) a second phase transition occurs from MCFL to the so-called
paramagnetic CFL phase (PCFL) \cite{Phases,PCFL}. This one is a real
phase transition because the PCFL phase breaks the translational
symmetry and the remaining rotational symmetry in the plane
perpendicular to the applied magnetic field. In the following
sections we will present the main characteristics of various CS
phases with either external or spontaneously generated magnetic
fields.

\section{Effect of strong magnetic fields on super-dense matter}

The symmetry breaking patterns of the MCFL and CFL phases are
different. In the CFL phase the symmetry breaking is given by
\begin{eqnarray}
\label{CFL} \mathcal{G}=SU(3)_C \times SU(3)_L \times SU(3)_R \times
U(1)_B \times U(1)_{\rm e.m.} \nonumber
 \\
\rightarrow SU(3)_{C+L+R}\times {\widetilde U(1)}_{\rm
e.m.}.\qquad\qquad\qquad\qquad\quad
\end{eqnarray}
This symmetry reduction leaves nine Goldstone bosons: a singlet
associated to the breaking of the baryon symmetry $U(1)_B$, and an
octet associated to the axial $SU(3)_A$ group. Once a magnetic field
is switched on, the difference between the electric charge of the
$u$ quark and that of the $d$ and $s$ quarks reduces the original
flavor symmetry of the theory and consequently also the symmetry
group remaining after the diquark condensate is formed. Then, the
breaking pattern for the MCFL-phase \cite{MCFL} becomes
\begin{eqnarray}
\label{MCFL} \mathcal{G_{B}}=SU(3)_C \times SU(2)_L \times SU(2)_R
\times U(1)^{(1)}_A\times U(1)_B \times U(1)_{\rm e.m.} \nonumber
 \\
\rightarrow SU(2)_{C+L+R} \times {\widetilde U(1)}_{\rm
e.m.}.\qquad\qquad\qquad\qquad\qquad\qquad\quad
\end{eqnarray}
The group $U(1)^{(1)}_A$ (not to be confused with the usual anomaly
$U(1)_{A}$) is related to the current which is an anomaly-free
linear combination of $s$, $d$, and $u$ axial currents
\cite{miransky-shovkovy-02}. In this case only five Goldstone bosons
remain. Three of them correspond to the breaking of $SU(2)_A$, one
to the breaking of $U(1)^{(1)}_A$, and one to the breaking of
$U(1)_B$. Thus, an applied magnetic field reduces the number of
Goldstone bosons in the superconducting phase, from nine to five.

The MCFL phase is not just characterized by a smaller number of
Goldstone fields, but also by the fact that all these bosons are
neutral with respect to the rotated electric charge. Hence, no
charged low-energy excitation can be produced in the MCFL phase.
This effect can be relevant for the low energy physics of a color
superconducting star's core and hence for its transport properties.
In particular, the cooling of a compact star is determined by the
particles with the lowest energy; so a star with a core of quark
matter and sufficiently large magnetic field can have a distinctive
cooling process.

Despite the difference between MCFL and CFL, the two phases are
hardly distinguishable at weak magnetic fields. The symmetry of the
CFL phase can be considered as an approximated symmetry in the
presence of an external magnetic field as long as the field-induced
mass of the charged Goldstone bosons is smaller than twice the gap,
so these mesons cannot decay into a quasiparticle-quasihole pair
\cite{Phases}. This explains why at a threshold field $\sim
\Delta^{2}_{CFL}$ a symmetry transmutation between the CFL and MCFL
phases takes place. Since, strictly speaking, the exact symmetry in
the presence of the magnetic field (ignoring quark masses) is that
of MCFL, the transition from the "approximated" CFL to MCFL at some
threshold field is not a phase transition, but a crossover or
symmetry transmutation. Only at fields comparable to
$\Delta^{2}_{CFL}$ the main features of MCFL emerge through the
low-energy behavior of the system. At the threshold magnetic field,
only five of the original nine Goldstone bosons that characterize
the low-energy behavior of the CFL phase remain. These are precisely
the five neutral Goldstone bosons determining the new low-energy
behavior of the genuinely realized MCFL phase.

As discussed in Ref. \cite{Phases} there is a close analogy between
the CFL-MCFL transmutation and what can be called a "field-induced"
Mott transition \cite{Mott}. Mott transitions have been discussed in
condensed matter \cite{Mott} and in QCD \cite{Mott-QCD}  to describe
delocalization of bound states into their constituents at a
temperature defined as the Mott temperature. By definition, the Mott
temperature $T_M$ is the temperature at which the mass of the bound
state equals the mass of its constituents, so the bound state
becomes a resonance at T > $T_M$. In the CFL-MCFL transmutation, the
usual role of the Mott temperature is played by the threshold
magnetic field.

Now, if we keep increasing the magnetic field until it reaches the
next energy scale $m_{M}\sim g\mu$, that is for $\widetilde{B} \geq
\widetilde{B}_{PCFL} = m_{M}^2$ \cite{PCFL}, one of the modes of the
charged gauge field becomes tachyonic. This phenomenon is a
consequence of the well known "zero-mode problem" for spin-1 charged
fields in the presence of a magnetic field found for Yang-Mills
fields \cite{zero-mode}, for the $W^{\pm}_{\mu}$ bosons in the
electroweak theory \cite{Skalozub}, and even for higher-spin fields
in the context of string theory \cite{porrati}. This effect is due
to the interaction of the applied magnetic field with the charged
gluon anomalous magnetic moment ($i\widetilde{e}\widetilde{f}_{\mu
\nu}G_{\mu}^{+}G_{\nu}^{-} $).

Similarly to other spin-1 theories with magnetic instabilities
\cite{zero-mode, Skalozub}, the solution of the zero-mode problem
leads to the restructuring of the ground state through the formation
of an inhomogeneous condensate $\overline{G}$ of the modulus of the
charged gluons, as well as the boost of the existing magnetic field
$\widetilde{\textbf{B}}=\nabla\times\widetilde{\textbf{A}}$ due to
the backreaction of the $\overline{G}$ condensate on the rotated
magnetic field. It can be corroborated \cite{PCFL} that the minimum
equation for the $\overline{G}$ field is equivalent to the
Ginzburg-Landau equation appearing in the Abrikosov's approach
\cite{Abrikosov} to type II metal superconductivity for the limit
situation when the applied field is near the critical value
$H_{c2}$. As in the Abrikosov's case, the order parameter,
$\overline{G}$, continuously increases from zero with the applied
field, thus signalizing a second-order phase transition where a
gluon crystalline vortex state with the corresponding magnetic flux
tubes condenses, thus breaking the translational symmetry and the
remaining rotational symmetry in the plane perpendicular to the
applied magnetic field.

A peculiarity of the present situation \cite{PCFL} is that contrary
to what occurs in conventional type-II superconductors, where the
applied field only penetrates through the flux tubes and with a
smaller strength, the gluon vortex state exhibits a paramagnetic
behavior; that is, outside the flux tube the applied field
$\widetilde{B}$ totally penetrates the sample, while inside the
tubes the magnetic field becomes larger than $\widetilde{B}$. Hence,
since the $\widetilde{Q}$ photons remain massless in the presence of
the condensate $\overline{G}$, the $\widetilde{U}(1)_{em}$ symmetry
remains unbroken. At asymptotically large densities, because
$\Delta_{CFL}\ll m_{M}$, we have $\widetilde{B}_{MCFL}\ll
\widetilde{B}_{PCFL}$ for each $\mu$ value.

Under the assumption that the contribution of the sextet (symmetric)
gaps can be neglected at all values of the magnetic field, a new
two-flavor phase, the $2SCds$, was recently found \cite{Warringa} at
fields much larger than $\mu^2$.

\section{Magnetic field generation in super-dense matter}

Matter inside compact stars should be neutral and remain in $\beta$
equilibrium. When these conditions along with the mass of the
s-quark, $M_s$, are taken into account in the dynamics of the
gluons, some gluon modes become tachyonic \cite{Huang,Fukushima},
indicating that the system ground state should be restructured. In
Ref. \cite{Vortex} we addressed this problem for the Meissner
unstable region of the so-called gapped 2SC phase.

As known, once the gluons are taken into consideration, the gapped
2SC becomes unstable at certain values of the baryon chemical
potential \cite{Huang}. As it turns out, the tachyonic modes in this
unstable phase are associated with the rotated charged gluon fields.
Note that even though the gluons are neutral with respect to the
regular electric charge, in the 2SC phase some gluons have nonzero
rotated charge. Since the charged gluons are the first becoming
tachyonic as the baryon chemical potential is decreased toward
moderate density values, it is natural to expect that the stable
ground state should incorporate the condensation of them. In Ref.
\cite{Vortex} we allowed for an inhomogeneous condensate
$\overline{G}$ of such gluons. Taking into account that this kind of
solution may generate rotated electromagnetic currents, the rotated
magnetic field $\widetilde{B}$ was also included in the general
framework of the condensation phenomenon, but now as an induced
field to be found self-consistently from the minimum equations of
the system free energy. The solution of the minimum equations
implied that the instability is actually removed by the formation of
a gluon condensate $\overline{G}$ that induces a magnetic field
$\widetilde{B}$. Since what condensates is the modulus of the
charged gluons, the condensate is neutral and thus preserves the
rotated electromagnetic gauge invariance $\widetilde{U}_{em}(1)$ of
the CS.

We underline that contrary to what occurs in the PCFL phase where a
strong applied magnetic field is needed to produce the inhomogeneous
gluon condensation, the condensation phenomenon here is connected to
a Meissner instability triggered by pairing stress. The pairing
stress develops at moderate densities due to the neutrality and
$\beta$-equilibrium of the medium. The spontaneous induction of a
rotated magnetic field in CS systems that have pairings with
mismatched Fermi surfaces is a new kind of phenomenon that can serve
to generate magnetic fields in stellar compact objects as magnetars.

When the absolute value of the magnetic mass becomes of order $m_g$,
the gluon condensate could produce a magnetic field of order
$10^{16}-10^{17}$ G. The possibility of generating a magnetic field
of such a large magnitude in the core of a compact star without
relying on a magneto-hydrodynamic effect can be an interesting
alternative to address the main criticism \cite{magnetar-criticism}
of the observational conundrum of the standard magnetar paradigm
\cite{magnetars}. Then, we conclude that if color superconductivity
is realized in the core of compact stellar objects at such expected
densities that a Meissner unstable phase is attained, the theory of
the origin of stellar magnetization should consider the mechanism
addressed in \cite{Vortex}. On the other hand, to have a mechanism
that associates the existence of high magnetic fields to CS at
moderate densities can serve to single out the magnetars as the most
probable astronomical objects for the realization of this high-dense
state of matter.

We stress that for even lower densities the contribution of
$\overline{G}$ and $\widetilde{B}$ in the quark quasiparticle
propagators cannot be neglected, thereby affecting the gap equation.
As a consequence, the inhomogeneity of the gluon condensate will be
naturally transferred to the diquark condensate. Hence, a LOFF-type
phase may appear as a back reaction of the inhomogeneous gluon
condensate on the gap solution. It is natural to expect that the
additional reduction in free-energy due to the
$\overline{G}-\widetilde{B}$ condensation will make this phase
energetically favored over a pure LOFF one \cite{LOFF}.

We anticipate that a $\overline{G}-\widetilde{B}$ condensate will
likely remove the chromomagnetic instability in the three-flavor
system too. In that case there are four gluons with tachyonic
masses. Following the results of the last paper in Ref.
\cite{Fukushima}, the four tachyonic gluons are $A_1$, $A_2$ and two
combinations of $A_3$, $A_8$ and $A_\gamma$. A third combination of
$A_3$, $A_8$ and $A_\gamma$ is massless, hence it represents the
rotated electromagnetic field in that phase, and the gluons $A_1$
and $A_2$ acquire rotated charge since they couple to the new
rotated electromagnetic field through its $A_3$ component. This
implies that $A_1$ and $A_2$ are analogous to the charged gluons
that become tachyonic in the 2SC case. If $A_1$ and $A_2$ condense
in an inhomogeneous condensate following the same mechanism we found
in Ref. \cite{Vortex}, this condensate could induce a rotated
magnetic field and also give real masses to the two combinations of
$A_3$, $A_8$ and $A_\gamma$ that were tachyonic. Exploring this idea
will require a clever approach that will permit us to circumvent the
difficulty of dealing with the big jump in the Meissner masses at
the onset of the chromomagnetic instability.

In addition to our mechanism \cite{Vortex} for spontaneous
generation of a magnetic field in a CS, a field may be also
spontaneously generated by an induced Goldstone boson current, as
reported in \cite{Son}.

\begin{theacknowledgments}
  We acknowledge the support of DOE Nuclear Theory
grant DE-FG02-07ER41458.
\end{theacknowledgments}





\begin{thebibliography}{9}

\bibitem{CS}D.~Bailin and A.~Love, \emph{Phys. Rep.} {\bf 107}, 325
(1984).

\bibitem{CFL} M. Alford, K. Rajagopal and F. Wilczek, \emph{Phys. Lett. B}
\textbf{422}, 247 (1998); \emph{Nucl. Phys. B} \textbf{537}, 443
(1999); R. Rapp, T. Schafer, E. V. Shuryak and M. Velkovsky,
\emph{Phys. Rev. Lett.} {\bf 81}, 53 (1998).

\bibitem{Taylor} I. Fushiki, E. H. Gudmundsson and C. J. Pethick, \emph{Astrophys.
J.} \textbf{342}, 958 (1989); T. A. Mihara, et al., \emph{Nature}
\textbf{346}, 250 (1990); G. Chanmugam, \emph{Ann. Rev. Astron.
Astrophys.} \textbf{30}, 143 (1992); J. H. Taylor, R. N. Manchester
and A. G. Lyne, \emph{Astrophys. J. S.} {\bf 88}, 529 (1993); D.
Lai, \emph{Rev. Mod. Phys.} {\bf 73}, 629 (2001).

\bibitem{magnetars} C.~Thompson and R.~C.~Duncan, \emph{Astrophys. J.} {\bf 392}, L9 (1992);
{\bf 473}, 322 (1996); S. Kulkarni and D. Frail, \emph{Nature} {\bf
365}, 33 (1993); T. Murakami et al., \emph{Nature} {\bf 368}, 127
(1994); Ibrahim et al., \emph{Astrophys. J.} {\bf 609}, L21 (2004).

\bibitem{Lai} L.~Dong and S.~L.~Shapiro, \emph{Astrophys. J.} {\bf 383}, 745 (1991).

\bibitem{alf-raj-wil-99/537}
M. Alford, K. Rajagopal and F. Wilczek, \emph{Nucl. Phys. B}
\textbf{537}, 443 (1999);M. Alford, J. Berges, and K. Rajagopal,
\emph{Nucl. Phys. B} \textbf{571}, 269 (2000).

\bibitem{MCFL}
  E.~J.~Ferrer, V.~de la Incera and C.~Manuel,
 \emph{Phys. Rev. Lett.} {\bf 95}, 152002 (2005); \emph{Nucl. Phys. B} {\bf
 747}, 88 (2006); \emph{PoS}  \emph{JHW2005} 022 (2006); \emph{J.\ Phys.\ A} {\bf 39}, 6349 (2006).

\bibitem{Phases}E.~J.~Ferrer and V.~de la Incera, \emph{Phys. Rev. D} \textbf{76}, 045011 (2007).

\bibitem{Warringa} K. Fukushima and H. J. Warringa,
\emph{Phys. Rev. Lett.} {\bf 100}, 032007 (2008).

\bibitem{Igor} J. L. Noronha and I. A. Shovkovy, \emph{Phys. Rev. D}
\textbf{76}, 105030 (2007).

\bibitem{PCFL} E.~J.~Ferrer and V.~de la Incera, \emph{Phys. Rev. Lett.}  {\bf 97}, 122301
(2006); \emph{J. Phys. A: Math. Theor.} {\bf 40}, 6913 (2007).

\bibitem{Vortex}
E.~J.~Ferrer and V.~de la Incera, \emph{Phys. Rev. D} \textbf{76},
114012 (2007); \emph{AIP Conf.Proc.} \textbf{947}, 401 (2007).

\bibitem{Son-2} D. T. Son, \emph{Phys. Rev. D} {\bf  59}, 094019 (1999).

\bibitem{miransky-shovkovy-02} V.~A.~Miransky, and I.~A.~Shovkovy,
\emph{Phys. Rev. D} {\bf 66}, 045006 (2002).

\bibitem{Mott} N.F. Mott, \emph{Metal-Insulator Transitions}, Taylor and Francis, London, 1974.

\bibitem{Mott-QCD} J. Hufner, S.P. Klevansky, and
P. Rehberg, \emph{Nucl. Phys. A} 606, 260 (1996).

\bibitem{zero-mode}V. V. Skalozub, \emph{Sov. J. Nucl. Phys.} \textbf{23},
113 (1978); N. K. Nielsen and P. Olesen, \emph{Nucl. Phys. B}
\textbf{144}, 376 (1978).

\bibitem{Skalozub}V. V. Skalozub, \emph{Sov. J. Nucl. Phys.} \textbf{43}, 665 (1986);
\textbf{45}, 1058 (1987). J. Ambjorn and P. Olesen, \emph{Nucl.
Phys. B} \textbf{315}, 606 (1989); \emph{Phys. Lett. B}
\textbf{218}, 67 (1989).

\bibitem{porrati}S. Ferrara and M. Porrati, \emph{Mod. Phys. Lett. A} \textbf{8},
2497 (1993); E. J. Ferrer and V. de la Incera, \emph{Int. Jour. of
Mod. Phys. A} \textbf{11}, 3875 (1996).

\bibitem{Abrikosov}A. A. Abrikosov, \emph{Sov. Phys. JEPT} \textbf{5}, 1174
(1957).

\bibitem{Huang} M. Huang and I. A. Shovkovy, \emph{Phys. Rev. D} {\bf 70}, 051501 (2004); {\bf 70}, 094030 (2004).

\bibitem{Fukushima} R.~Casalbuoni, et. al. \emph{Phys. Lett. B} \textbf{605}, 362 (2005);  \textbf{615}, 297(E)
(2005); M. Alford and Q. H. Wang, \emph{J. Phys. G} \textbf{31}, 719
(2005); K. Fukushima, \emph{Phys. Rev. D} \textbf{72}, 074002
(2005).

\bibitem{magnetar-criticism}
J. Vink and L. Kuiper, \textit{Mon.\ Not.\ Roy.\ Astron.\ Soc.\
Lett.}  {\bf 370}, L14 (2006).

\bibitem{LOFF} M. Alford, J. A. Bowers, and K. Rajagopal, \emph{Phys. Rev. D} \textbf{63},
074016 (2001); I. Giannakis and H.C. Ren, \emph{Phys. Lett. B}
\textbf{611}, 137 (2005); \emph{Nucl. Phys. B} \textbf{723}, 255
(2005); R.~Casalbuoni, et.al, \emph{Phys. Lett. B} \textbf{627}, 89
(2005); M. Mannarelli, K. Rajagopal and R. Sharma, \emph{Phys. Rev.
D} \textbf{73},114012 (2006); M. Ciminale, G.~Nardulli, M. Ruggieri
and R. Gatto, \emph{Phys. Lett. B} \textbf{636}, 317 (2006); K.
Rajagopal and R. Sharma, \emph{Phys. Rev. D} \textbf{74}, 094019
(2006).


\bibitem{Son}D.~T. Son, M.~A. Stephanov, \emph{Phys. Rev. D} \textbf{77}, 014021 (2008).
\end{thebibliography}
\end{document}